\documentclass[]{spie} 
\pdfoutput=1 
\usepackage[]{graphicx}
\usepackage{subcaption}
\usepackage{xcolor}

\usepackage[utf8]{inputenc}
\usepackage[T1]{fontenc}
\usepackage{parskip}
\usepackage{wasysym}

\usepackage{graphicx}
\usepackage{media9}
\usepackage{cite}
\usepackage{hyperref}

\title{Performance of Large-Format Deformable Mirrors Constructed with TNO Variable Reluctance Actuators }

\author{Rachel Bowens-Rubin\supit{a}, Philip Hinz\supit{a}, Wouter Jonker\supit{b}, 
Stefan Kuiper\supit{b}, Cesar Laguna\supit{a}, and Matthew Maniscalco\supit{b}
\skiplinehalf
\supit{a}UC Santa Cruz, 1156 High St, Santa Cruz CA, USA; \\
\supit{b} TNO Technical Sciences, Delft, The Netherlands
}
\authorinfo{Further author information: (Send correspondence to R. Bowens-Rubin)\\R.B.R.: E-mail: rbowru@ucsc.edu}

\begin{document} 
\maketitle 
  
  \begin{abstract}
Advancements in making high-efficiency actuators are an enabling technology for building the next generation of large-format deformable mirrors. The Netherlands Organization for Applied Scientific Research (TNO) has developed a new style of variable-reluctance actuator that requires approximately eighty times less power to operate as compared to the traditional style of voice-coil actuators.  We present the performance results from laboratory testing of TNO's 57-actuator large-format deformable mirror from measuring the influence functions, linearity, hysteresis, natural shape flattening, actuator cross-coupling, creep, repeatability, and actuator lifetime.   We measure a linearity of 99.4 $\pm$ 0.33\% and hysteresis of 2.10 $\pm$ 0.23\% over a stroke of 10 microns, indicating that this technology has strong potential for use in on-sky adaptive secondary mirrors (ASMs).  We summarize plans for future lab prototypes and ASMs that will further demonstrate this technology.  

\end{abstract}

 
\section{INTRODUCTION}

\indent \indent Adaptive secondary mirrors have been used successfully at telescopes like the Large Binocular Telescope~\cite{Esposito2010}, Magellan~\cite{Close2018}, the VLT Observatory~\cite{Biasi2012}, and the MMT Observatory~\cite{Wildi2003}. However, these adaptive secondary mirror systems are complex  because they incorporate hundreds of power intensive voice-coil actuators that require a dedicated cooling system and capacitive sensors~\cite{Hinz2016}. The mirror facesheet fabrication has also traditionally required the grinding of a large piece of glass into a thin glass sheet that is flexible enough to be shaped by the actuators. 

The Netherlands Organization for Applied Scientific Research (TNO) has made significant breakthroughs in large-format deformable mirror technology that could enable adaptive secondary mirrors to become simpler and less costly.   The key advancement is a new style of hybrid variable reluctance actuator that is 80 times more efficient than traditional voice-coil actuators (TNO actuator efficiency~\cite{Kuiper2018} = $40 N/\sqrt{W}$; MMT and LBT actuator efficiency~\cite{Riccardi2003} = $0.5 N/\sqrt{W}$). The low required power offers a pathway to eliminate the actuator cooling system, miniaturize the system, and pack the actuators at higher density. TNO is also developing solutions that could simplify the facesheet fabrication by starting with thin glass sheets and utilizing glass-slumping techniques to form the shape.  

While these mirrors show great potential to make advanced adaptive optics systems accessible to a wide variety of observatories, they have yet to be proven to the astronomical community. % significant potential to a
This paper presents laboratory testing performed of TNO's third deformable-mirror prototype (DM3) to explore the properties of TNO's large-format deformable mirror technology. We report the results of this testing and outline the future steps towards deployment of this technology on-sky.

\section{TECHNOLOGY OVERVIEW}
\subsection{Design of the TNO DM3}

The DM3 is the third prototype large-format deformable mirror from TNO  (Figure 1). It was built for lab demonstration in 2016. The surface is round (\diameter 150mm) and  flat to within 15.5 um peak-to-valley when the applied actuator current is zero.  DM3 was built using 57 actuators of 18mm pitch (Figure \ref{fig:DM3acts}).  

The mirror is controlled through a 64 analog output from a Real Time Linux environment. The output voltages are controlled using an RDA interface to Matlab running on an office PC. The user can control the shape of the mirror's surface by setting the current applied to each actuator. The update rate of this setup is 300Hz.  

\begin{figure} [h!]
  \centering
  {\includegraphics[width=0.5\textwidth]{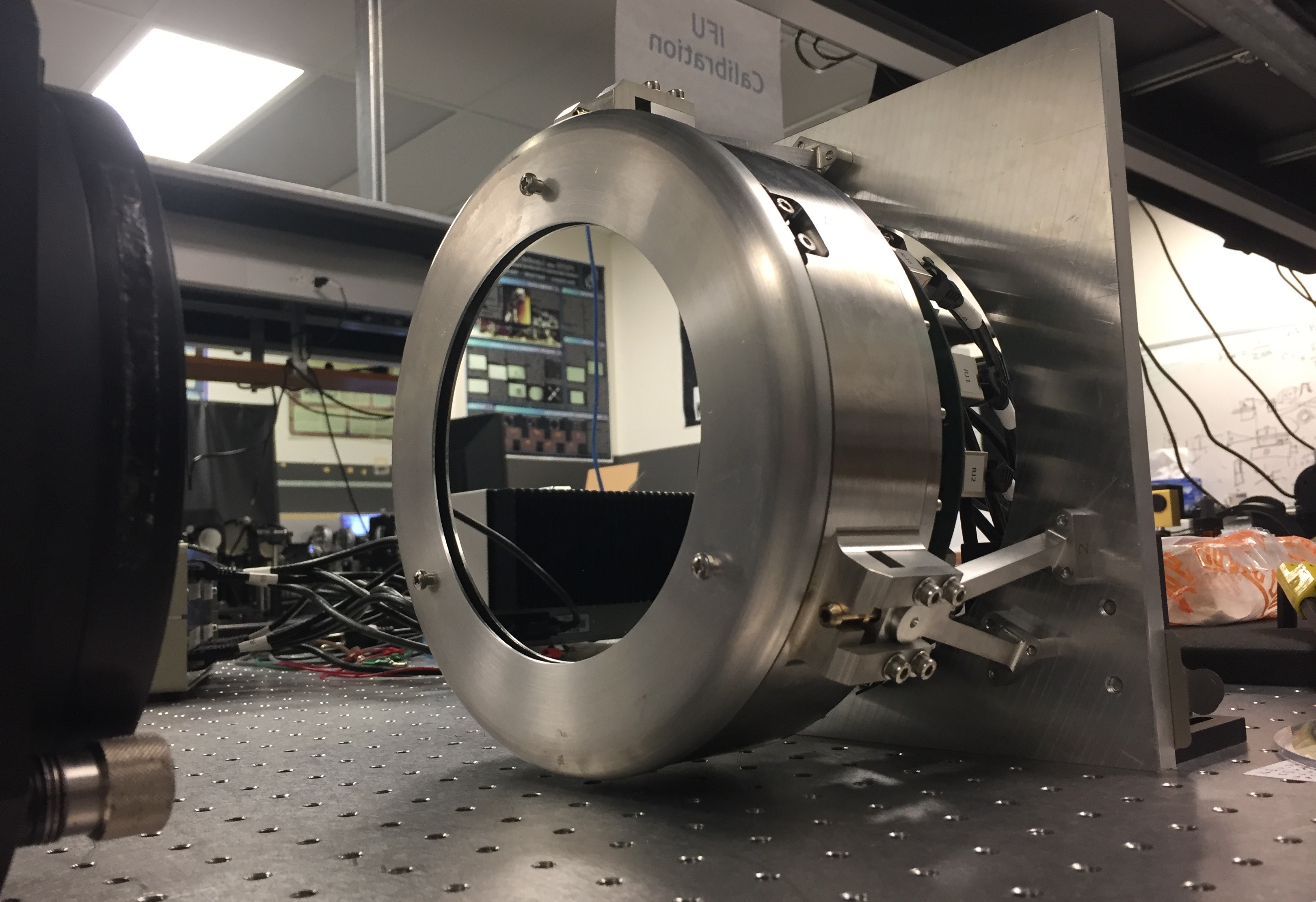}}~                
  {\includegraphics[width=0.45\textwidth]{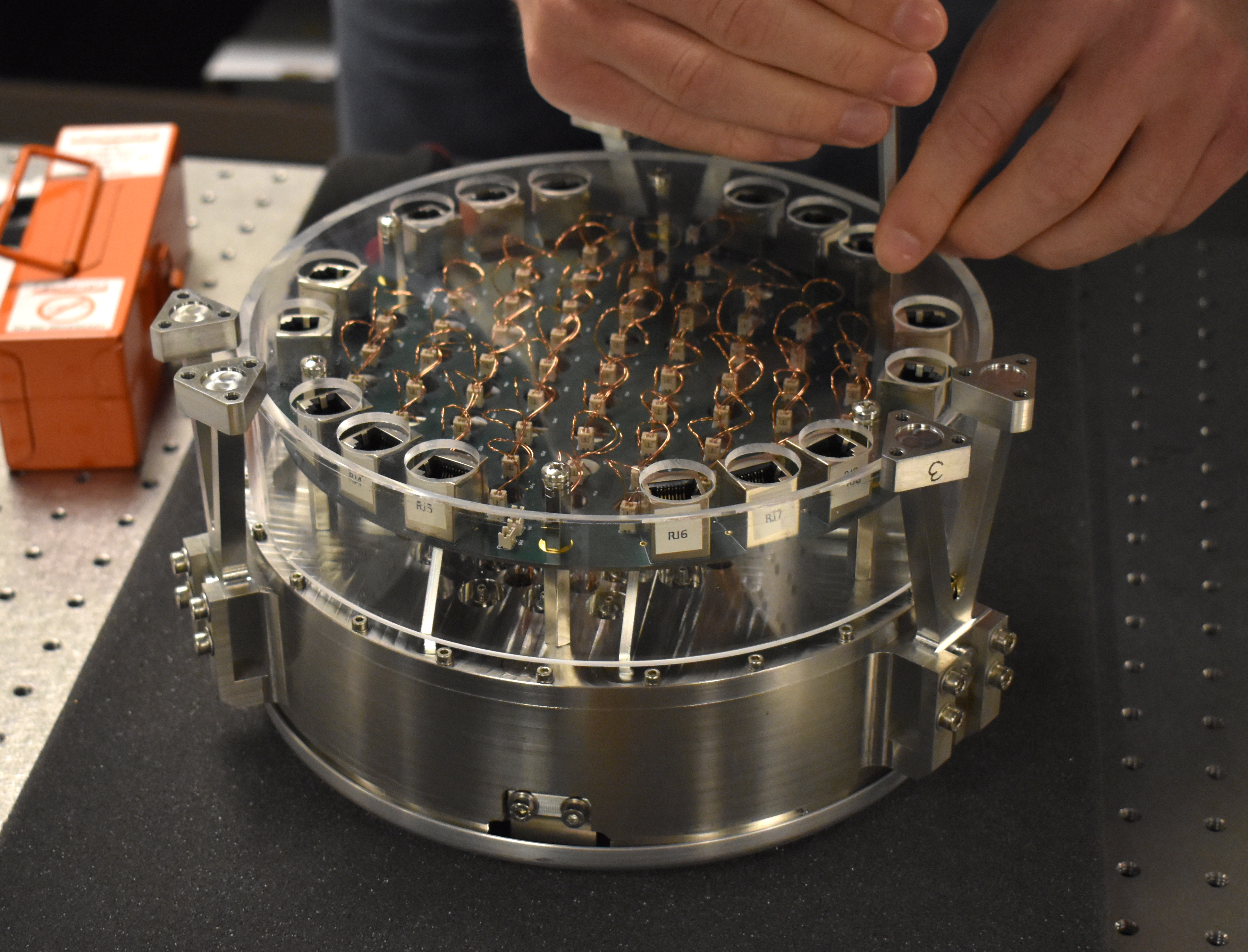}}~
 \caption{ \textbf{DM3 at the UCSC Lab for Adaptive Optics}. The third prototype large-format deformable mirror from TNO (DM3) contains 57 variable reluctance actuators.  }
     \label{tnoinlao}
\end{figure}

\begin{figure} [h!]
  \centering
  {\includegraphics[width=0.45\textwidth]{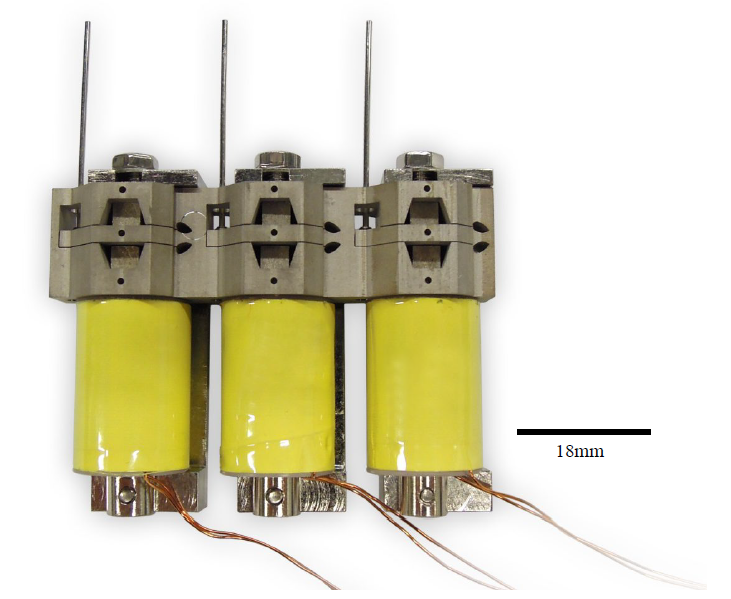}}~                
  {\includegraphics[width=0.45\textwidth]{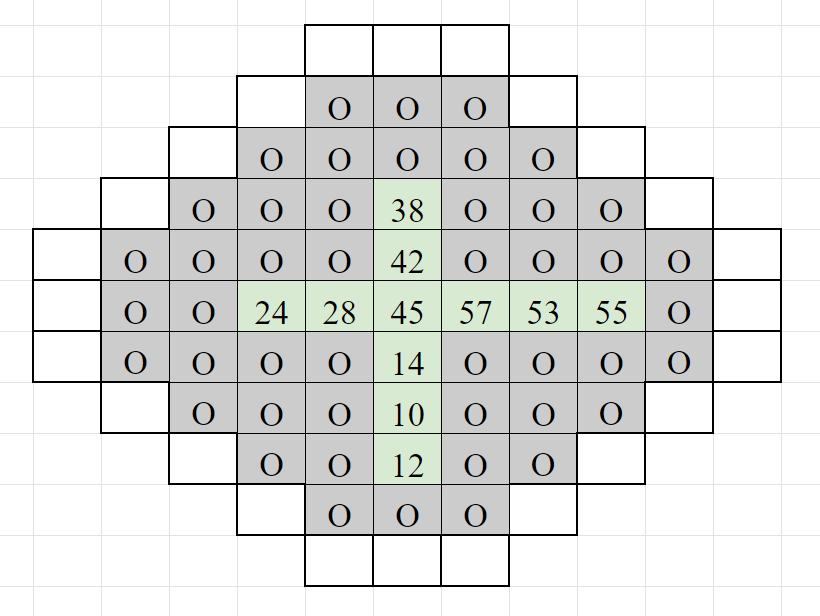}}~
 \caption{\textbf{TNO Variable Reluctance Actuators} (Left) \emph{The 18mm-pitch TNO Actuator used in DM3.} (Right) \emph{Actuator layout in DM3.} The actuators used for the testing reported in this paper are labeled by number in the green boxes. The other actuators are shown in the gray boxes, and the channels unoccupied are shown in white.}
     \label{fig:DM3acts}
\end{figure}

\subsection{TESTBENCH SETUP}
\begin{figure}[h!]
   \begin{center}
   \begin{tabular}{c}
   \includegraphics[height=7cm]{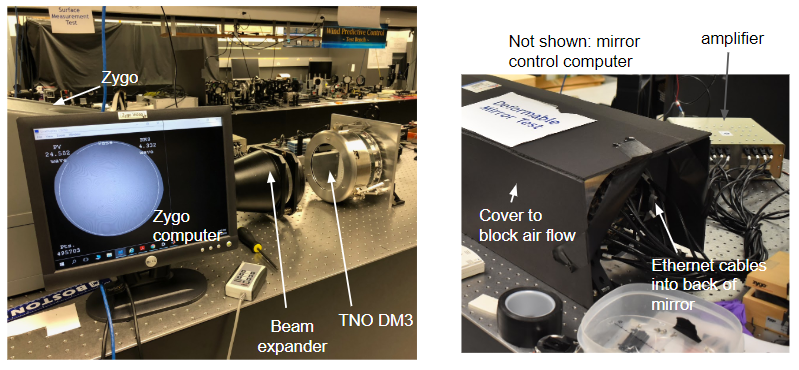}
   \end{tabular}
   \end{center}
   \caption[example] 
   { \label{fig:testbed} 
\textbf{Large-Format Deformable Mirror Testbench at the UCSC Lab for Adaptive Optics}. The testbench utilizes a Zygo Interferometer with MetroPro software to measure surface shape. A beam expander is used so that the majority of the DM3 can be measured. }% TNO DM3 mirror.}
   \end{figure}

The University of California Santa Cruz Lab for Adaptive Optics (UCSC-LAO) began testing the TNO DM3 in January of 2020. The testbed at UCSC-LAO (Figure \ref{fig:testbed}) utilizes a Zygo Interferometer run with Metropro Software to measure the shape of the mirror’s surface to a precision of 0.6 nm. 
Additional Matlab code is being developed by UCSC-LAO to streamline the process of characterizing the actuator linearity, hysteresis, creep, cross-coupling, repeatability, Zernike pattern replication, and lifetime using this testbench. A cold chamber and orientation setup will be integrated for future environmental and gravity testing.

\newpage

\section{DM3 TESTING RESULTS}

\subsection{Influence Functions}
The influence function of each actuator was measured using an individual actuator poke test.  A $+20mA$ current was applied in an automated measurement sequence to displace each actuator one-by-one. The displacement of these pokes corresponded to approximately 730nm peak-to-valley.   Twenty-one actuators responded to the positive current with a positive  displacement, and thirty-six actuators responded with a negative displacement.   

\begin{figure}[ht]
   \begin{center}
   \begin{tabular}{c}
   \includegraphics[width=1.0\textwidth]{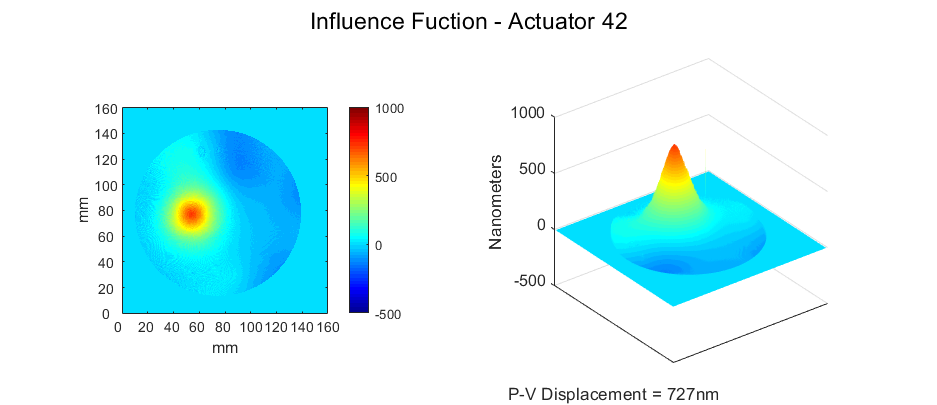}
   \end{tabular}
   \end{center}
   \caption[example] 
   { \label{fig:exampleinflufun} 
   \textbf{Example Influence Function}: The influence function for Actuator 42 as measured on April 30th, 2020. An applied voltage of +20mA resulted in an actuator displacement of 727nm.  Images are scaled with the x-y axis in millimeters and the color axis and z-displacement in nanometers.     }
   \end{figure}

\begin{figure}[h!]
   \begin{center}
   \begin{tabular}{c}
   \includegraphics[width=0.6\textwidth]{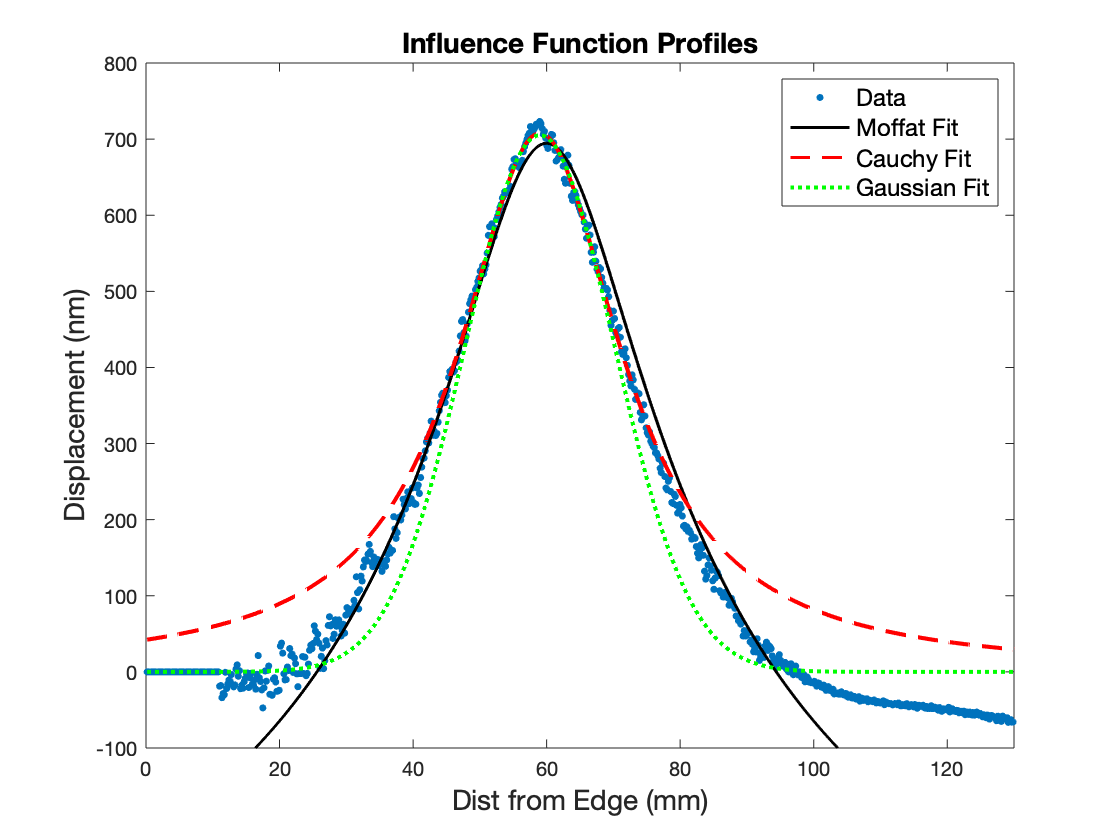}
   \end{tabular}
   \end{center}
   \caption[example] 
    { \label{fig:profile} {\textbf{Influence Function Profile}}. A cross-section of the influence function for Actuator 42 was used to determine a profile of best fit. A Moffat,  Cauchy, and Gaussian curve were fit to the inner 100 points of the cross section. The Cauchy and Gaussian fit approximate the center well, but do not approximate the width correctly.  The Moffat fit can approximate the width to higher precision, but cannot be used to determine the center value. A custom profile is needed to match the greater shape of the cross section.  }
   \end{figure}

An example of the influence function measured for Actuator 42 is shown in Figure \ref{fig:exampleinflufun}. The influence function was measured with the natural shape of the mirror subtracted such that only the displacement is seen.  The cross-section of the influence function is plotted in Figure \ref{fig:profile} with a Gaussian, Moffat, and Cauchy function fit. While each of the three functions can approximate part of the profile, a custom profile would need to be created to correctly fit the actuator cross sections at both the center and tails. 

%\clearpage
\subsection{Natural-Shape Surface Flattening} 

The natural shape of the TNO DM3 mirror has a peak-to-valley that is 15.5 um with a 2780nm RMS variation (within $\diameter$140mm). To bring the surface as flat as possible, a set of currents that could bring the mirror surface to the lowest RMS possible was determined using an iterative process.  At each step, an image was taken using the Zygo. The next desired position for the actuators to reproduce in the iteration ($i$) was set to the negative of the last image taken ($S_i$). % and multiplied by -1 to create a "desired shape" image.   
The next set of currents ($F_i$) to produce the desired shape ($S_i$) were calculated using the influence functions matrix ($P$) through the  Moore-Penrose inverse: 

\begin{equation}
  F_i = (P^{T}P)^{-1} P^T S_i.
\end{equation}

\noindent The new currents from the iteration were then summed with the previous currents applied to find the new values to be applied ($F = \sum_{1}^{i = n} F_i$). 

\begin{figure}[ht]
   \begin{center}
   \begin{tabular}{c}
   \includegraphics[width=0.9\textwidth]{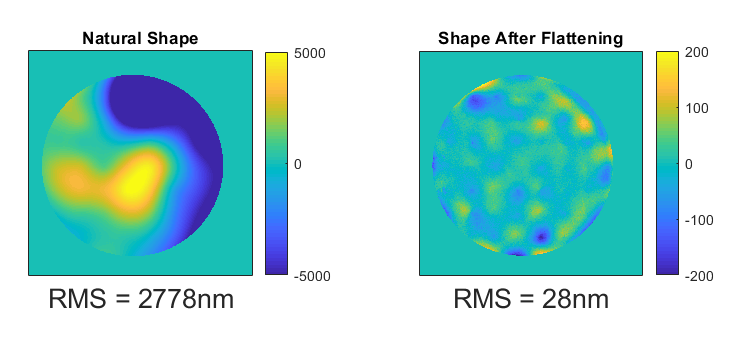}
   \end{tabular}
   \end{center}
   \caption[example] 
%>>>> use \label inside caption to get Fig. number with \ref{}
   { \label{fig:flat} 
\textbf{Natural-Shape Surface Flattening}. (Left) \emph{DM3 Natural Surface Shape}. Unpowered, the shape of the TNO DM3 mirror surface has a peak-to-valley that is 15500nm (RMS = 2780nm within \diameter140mm).  (Right) \emph{Surface Shape after Applied Flattening}.
  After flattening, the surface shape was brought down to 28nm RMS with a peak-to-valley of 469 nm. The colorbar is reported in units of nanometers.
 }
   \end{figure}

 On April 30th 2020, fifteen iterations of flattening were performed to calculate the optimal flattening currents.  The majority of the variation was removed after five iterations. The flattest shape was found on iteration 14 which had an RMS = 27.7nm (within $\diameter$140mm) with a peak-to-valley of 469nm, demonstrating that it is possible to flatten the surface by two orders of magnitude. Figure \ref{fig:flat} shows the natural shape of the mirror alongside the shape after this flattening. 
 
 The total current needed to hold this shape across the 57 actuators was 1967mA, averaging 34.5 $\pm$ 6.6mA per actuator. The required power was 189mW (3.3 mW/actuator).  Riccardi et al. (2003, Table 1) states the power required for a correction to a similar wavefront error with the LBT ASM requires 190 mW/actuator.~\cite{Riccardi2003}, leading to a nearly sixty factor gain in this comparison.   This estimate does not include the additional 3.8 W/actuator of power needed at the LBT collocated electronics crate, which is located slightly further from the mirror but still on the secondary structure.  

\subsection{Zernike Mode Testing} 

\begin{figure}[ht]
   \begin{center}
   \begin{tabular}{c}
   \includegraphics[width=0.8\textwidth]{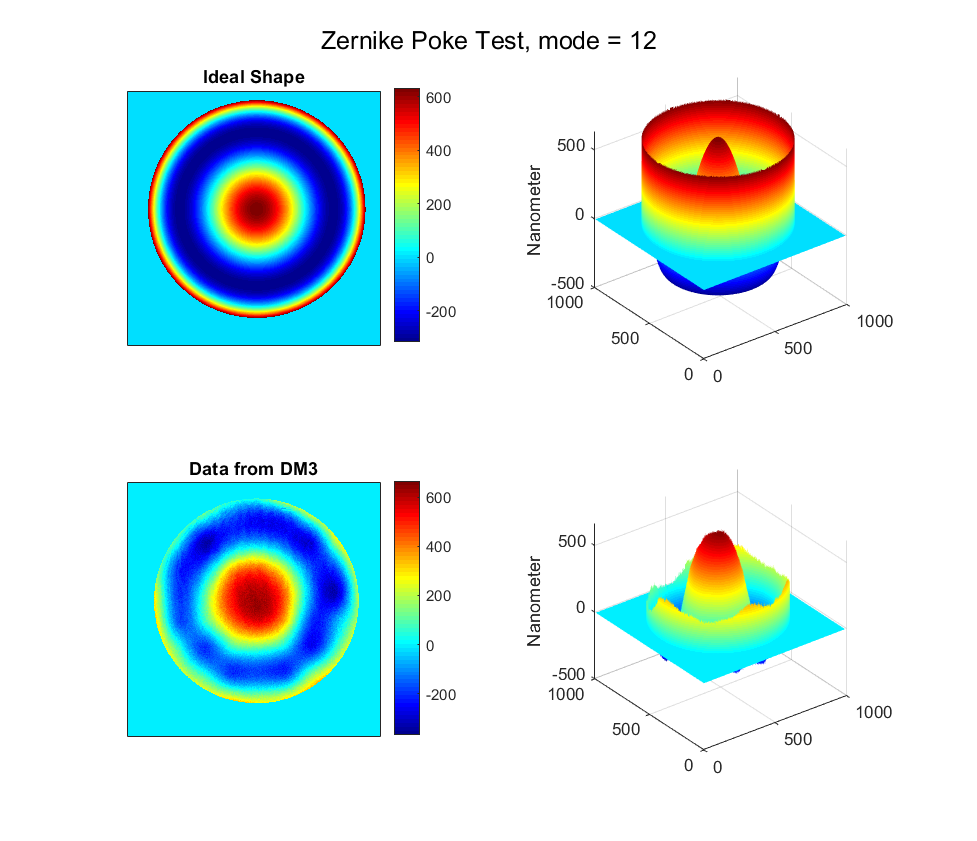}
   \end{tabular}
   \end{center}
   \caption[example] 
   { \label{fig:zernshape} 
\textbf{Example Zernike Mode}: Zernike Mode 12.  The ideal shape (top) is presented with the approximated shape made by  DM3 (bottom).  A movie of the full Zernike test data from applying modes 1 - 57 to DM3 can be found at the following youtube link: \href{https://youtu.be/dpTzO46zg2Y}{{\color{blue}https://youtu.be/dpTzO46zg2Y}}.   
 }
    \end{figure}

To determine how close DM3 was able reproduce ideal Zernike shapes,  the first 57 Zernike Modes were applied to DM3.  The currents needed to reproduce the Zernike modes were calculated using Equation 1, where $S_i$ was the shape of each Zernike pattern. No iterations were used in trying to reproduce the Zernike pattern. An example of the side-by-side comparison between DM3 and an ideal Zernike pattern can be viewed in Figure \ref{fig:zernshape} for mode 12.  A movie of the full test can be watched at the following youtube link: \href{https://youtu.be/dpTzO46zg2Y}{{\color{blue}https://youtu.be/dpTzO46zg2Y}}. 

\begin{figure}[ht]
   \begin{center}
   \begin{tabular}{c}
   \includegraphics[width=0.6\textwidth]{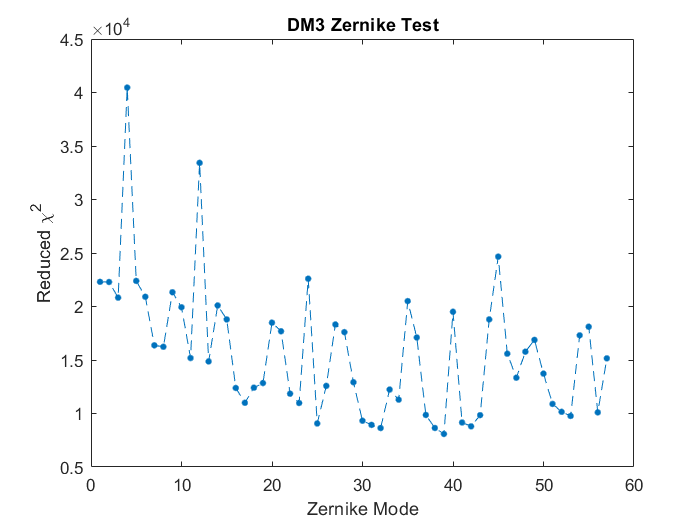}
   \end{tabular}
   \end{center}
   \caption[example] 
   { \label{fig:zernchisquare} 
\textbf{Zernike Mode Test Reduced Chi Square}.  The reduced chi square of each Zernike data frame was measured to determine how well each Zernike mode could be reproduced. Four of the spikes in the reduced chi square plot corresponded to the spherical Zernike modes (4, 12, 24, 40), indicating that these shapes were not as precisely replicated as the non-spherical modes.    
 }
   \end{figure}

To determine if a trend existed between the Zernike order number and how well the mirror could replicate the pattern, the reduced chi square was calculated for each Zernike frame (Figure \ref{fig:zernchisquare}).  Modes with a large reduced chi square value could not be replicated to as high precision. Overall, the spherical modes were reproduced the least well (4, 12, 24, 40) of any categorization of mode type. There was not a significant trend between the value of the reduced chi square and low or high order mode number. 

\subsection{Actuator Cross-Coupling}

\begin{figure}[h!]
   \begin{center}
   \begin{tabular}{c}
   \includegraphics[width=6in]{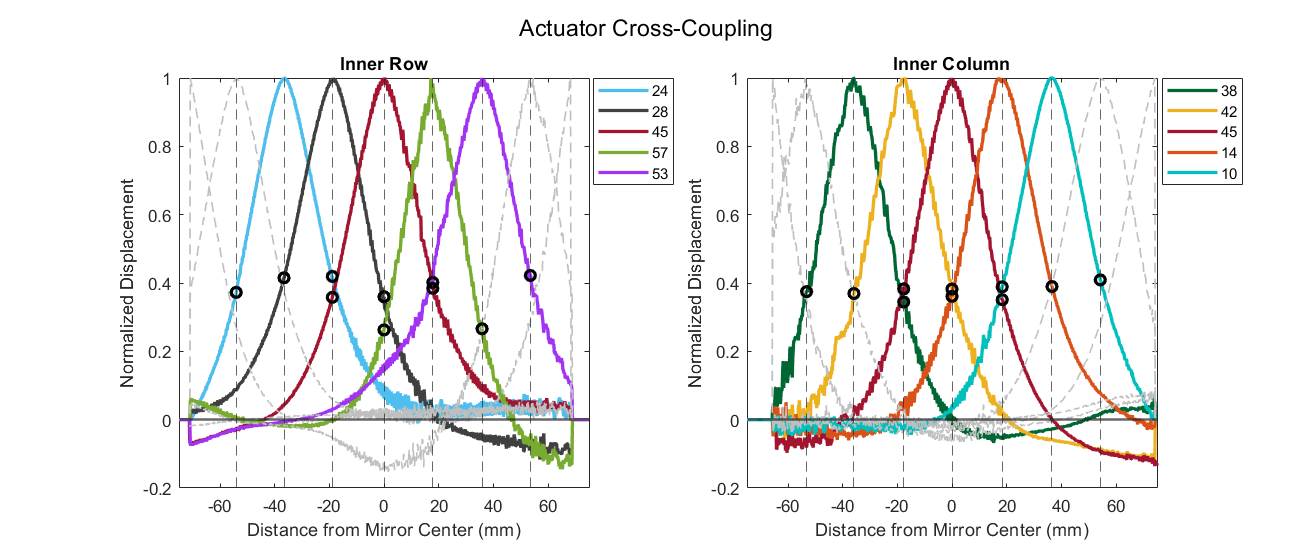}
   \end{tabular}
   \end{center}
   \caption[example] 
%>>>> use \label inside caption to get Fig. number with \ref{}
   { \label{fig:crosssection} 
\textbf{Actuator Cross-Coupling}. Five actuators from the inner row (24, 28, 45, 57, and 53) and five actuators from the inner column (38, 42, 45, 14, and 10) were used to measure the actuator cross-coupling. The filled lines represent the normalized cross section of the actuators included in the actuator cross-coupling measurement.  The black dots indicate the values of the cross-coupling measurements. The dashed lines are the influence functions of the actuators in the row or column not used for the cross-coupling calculation. } 
   \end{figure}

The actuator cross-coupling was measured to quantify how the movement of one actuator will affect its neighbor. The cross-coupling was measured using the cross section of the influence functions from the center row and center column of actuators (Figure \ref{fig:crosssection}).  The center actuator (45) was used in both samples. 

The cross sections were corrected for their positive/negative displacement polarity and then normalized. Two measurements were taken per actuator along the line that intersected the neighboring actuator's peak. Using these twenty measurements, we measure an actuator cross-coupling of 37.07 $\pm$  0.93\%. 

\subsection{Linearity} 

\begin{figure}[ht]
\centering
\begin{subfigure}{0.49\textwidth}
  \centering
  \includegraphics[width=1.0\linewidth]{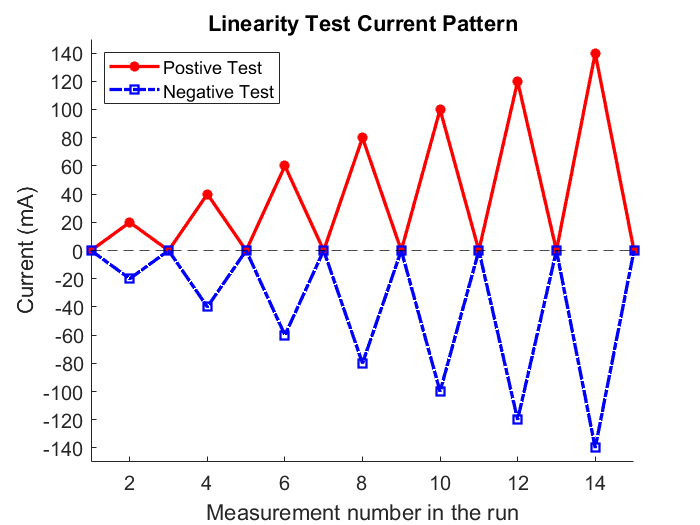}
 % \caption{\textbf{Linearity}}
  \label{fig:lincurrentpattern}
\end{subfigure}
\begin{subfigure}{.49\textwidth}
  \centering
  \includegraphics[width=1.0\linewidth]{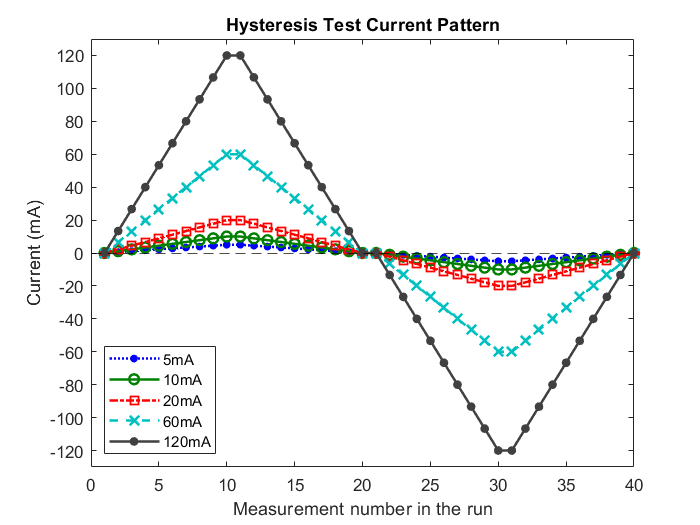}
 % \caption{\textbf{Hysteresis}}
  \label{fig:hystcurrentpattern}
\end{subfigure}
 \caption[example]{\textbf{Current Patterns Used in Testing}. (Left)  Linearity testing was performed in fifteen steps with a positive and negative current pattern. The  positive and negative current runs were combined by actuator for linearity fitting. (Right) Hysteresis runs were performed using forty steps across five maximum current values. Each run was analysed separately. }
 \label{fig:CurrentPatterns}
   \end{figure}

Linearity testing was performed using five actuators (12, 14, 42, 55, and 57). These actuators were run through a fifteen-step current pattern using positive and negative currents, spanning from 0 to +140mA or 0 to -140mA (Figure \ref{fig:CurrentPatterns}).  Although the negative and positive current runs were performed separately, linearity fitting considered the tests jointly for each actuator.  
The displacement was measured using the mean of the seven points nearest to the actuator center.  

\begin{figure}[ht]
   \begin{center}
   \begin{tabular}{c}
   \includegraphics[width=7in]{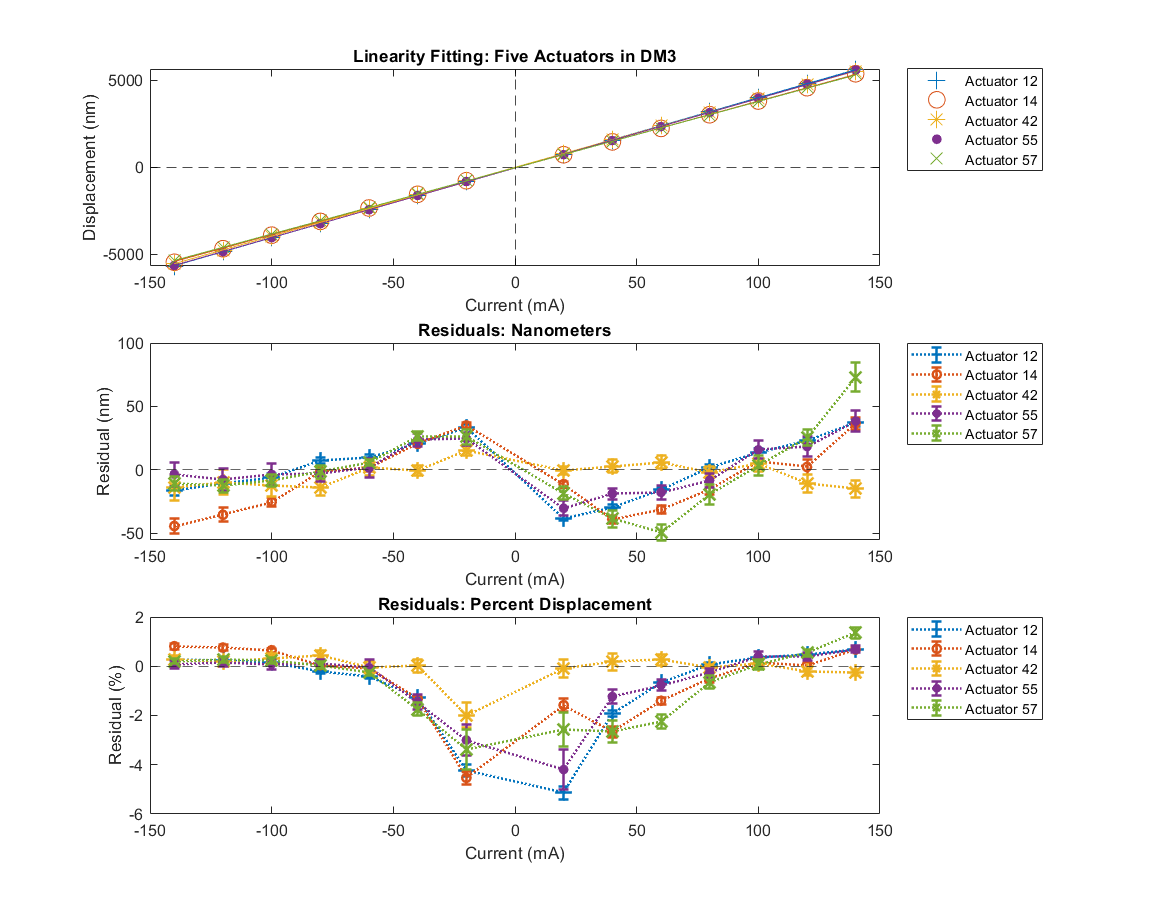}
   \end{tabular}
   \end{center}
   \caption[example] 
%>>>> use \label inside caption to get Fig. number with \ref{}
   { \label{fig:linearityresults} \textbf{Linearity Results.}
Linearity testing was performed using five actuators. Actuators 12, 14, 55, and 57 were wired with negative polarity but are adjusted here to be corrected to positive polarity.  The average actuator displacement was 39.66 $\pm$ 0.54 nm/mA. Displacement was linear to within 99.4\% $\pm$ 0.33\%.  A negative percent displacement occurs at small current values, meaning the actuator undershot the expected position. }
   \end{figure}

The displacement data were fit for each actuator (Figure \ref{fig:linearityresults}). The slopes of the fit for each actuator were averaged to reveal a displacement of 39.66 $\pm$ 0.54 nm/mA.  
The linearity data for the five actuators is presented in Figure \ref{fig:linearityresults} with the percent residuals.  The percent residuals were calculated using,

\begin{equation}
    Residual(\%) = 100 \left(\frac{Data  - Fit}{Data}\right). 
\end{equation}

The mean of the percent residuals across the five actuators was used to quantify the actuator linearity. We measure the actuators to be linear to  99.4\% $\pm$ 0.33\% (nonlinear to 0.6\% $\pm$ 0.33\%).  The nonlinearity is most evident at small currents (within $\pm$20mA) where the actuators tend to undershoot their predicted position.

% ===============

% ==================

\newpage
\subsection{Hysteresis}

Actuator displacement varies depending on the previous position of the actuator.  To quantify this, we measured the hysteresis using the definition: 

\begin{equation}
    Hyst(\%) = 100 \left(\frac{|S3-S1|}{|S2-S4|}\right)
\end{equation}

\noindent \noindent where $S1$ and $S3$ are the displacements measured at an applied current of zero and $S2$ and $S4$ are the displacements measured at the maximum and minimum current applied. 

\begin{figure}[ht]
\centering
\begin{subfigure}{1.0\textwidth}
  \centering
  \includegraphics[width=0.85\linewidth]{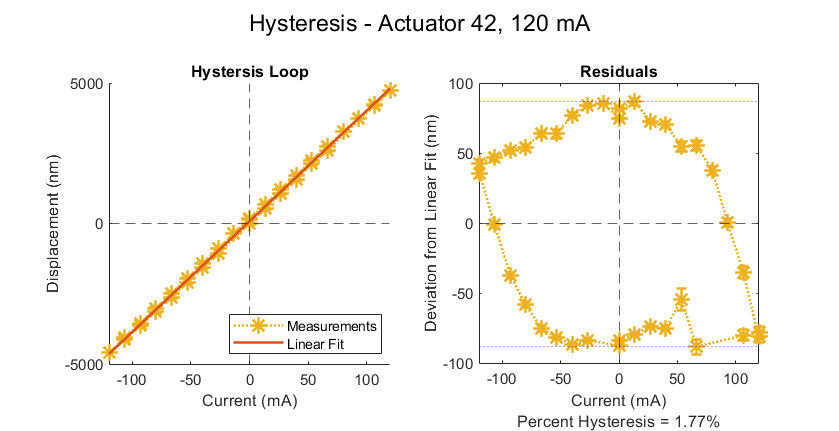}
 % \caption{\textbf{Example Hysteresis Loop}}
  \label{fig:hystcurrentpattern}
\end{subfigure}

\begin{subfigure}{1.0\textwidth}
  \centering
   \includegraphics[width=0.65\linewidth]{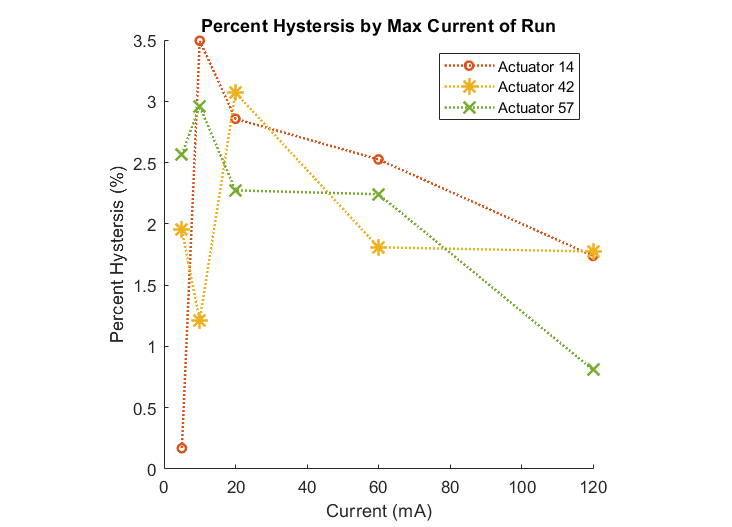}
 \end{subfigure}
 \caption[example]{\textbf{Hysteresis testing results}. (Top) A linear fit was performed and the residuals were plotted for each hysteresis run. The example for Actuator 42 for the 120mA run is shown here.   (Bottom) The percent hysteresis was measured in fifteen trials to be 2.10 $\pm$ 0.23\%.  }
 \label{fig:hyst}
   \end{figure}

Hysteresis testing was performed using three actuators (14, 42, and 57) stepped through five independent hysteresis loops. Each loop was done in forty steps between currents of $\pm$5mA,  $\pm$10mA, $\pm$20mA, $\pm$60mA, $\pm$120mA (Figure \ref{fig:CurrentPatterns}). Three of these tests occurred in the least linear portions of the actuator displacement curves ($\pm$5mA,  $\pm$10mA, $\pm$20mA) and provide  worse-case scenario measurements of the hysteresis.  No measurement exceeded a hysteresis value of 3.5\%.  
The average percent hysteresis measured across all fifteen trials was 2.10 $\pm$ 0.23\% (Figure \ref{fig:hyst}). 
   
\subsection{Repeatability}

\indent \indent A repeatability test was run to verify the ability for the DM3 to return to the same shape when applying the same set of currents from day to day.  The repeatability testing was performed by applying the flattening current set calculated on April 30th 2020.
Five measurements were taken across five days starting from zero currents applied. Images were visually inspected and the RMS was used to quantify the shape. The images passed visual inspection and the RMS did not change to within the  precision of the Zygo Interferometer (0.6nm). 

The same shape was applied after the mirror had been running for unmeasured amounts of times. The RMS differences in these images were observed to fluctuate by up to $\Delta RMS = 33.5$ nm. 

\subsection{Creep}
Creep testing was performed by applying a current pattern and holding it for twelve hours. Similar to the repeatability testing, the creep testing was run by applying the flattening current pattern found on April 30th 2020.  Measurements were taken with the applied pattern subtracted such that the Zygo images were more sensitive to any deviations. The tests were repeated three times.

The only measurable change came from an added tilt on the order of 1000nm that was believed to be from a shift in the beam expander mounting.  Although the data cannot conclusively rule out creep as an issue to  <1000nm RMS, there is no evidence for significant creep.

\subsection{Lifetime Testing}

A demonstration to test the working lifetime of the actuators is in progress. 
 The 57 actuators were set to a 34Hz cycle
going from -50\% to +50\% of the maximum stroke. This scenario is a strong worst case for an actual ASM in operation, where the changes in deflection for each cycle will be much lower. Periodically, DM3 is imaged with the Zygo interferometer. At the time of writing, each of the 57 actuators has seen over
110 million cycles with no detectable change in performance.

\clearpage
\section{CONCLUSIONS}
\subsection{Summary of Results}
The key results from each test of the TNO DM3 are summarized in Table 1. 
The values determined from the laboratory testing at UCSC-LAO of the TNO DM3 are consistent with the values quoted by TNO. The TNO variable reluctance actuators show a linearity and hysteresis value similar to the voice-coil actuators currently used in on-sky adaptive secondary mirrors while exhibiting over 5.5 times lower power consumption.  

\begin{table}[h!]
\center
\caption{\textbf{Summary of DM3 Testing Results}}
\begin{tabular}{|c|c|l|}
\hline
\textbf{Section} & \textbf{Test}                                                               & \multicolumn{1}{c|}{\textbf{Takeaway}}                                                                                                                                                                                                    \\ \hline
3.1              & Influence functions                                                         & \begin{tabular}[c]{@{}l@{}}The influence function center location can be approximated\\  with a Gaussian or Cauchy.  The width  can be approximated \\ using a Moffit fit.\end{tabular}                                                        \\ \hline
3.2              & \begin{tabular}[c]{@{}c@{}}Natural Shape \\ Surface Flattening\end{tabular} & \begin{tabular}[c]{@{}l@{}}The surface was brought from RMS = 2778nm to RMS = 28nm \\ with an average current per actuator = 34.5 $\pm$ 6.6mA, \\ total power = 189mW.\end{tabular}                                                    \\ \hline
3.3              & Zernike Mode                                                                & \begin{tabular}[c]{@{}l@{}} There is not a significant trend between mode number and pattern \\ replication. The spherical modes were the least well replicated. \end{tabular}                                                                                                                          \\ \hline
3.4              & Actuator Cross-Coupling                                                     & Cross Coupling = 37.07 $\pm$  0.93\%                                                                                                                                                                                                      \\ \hline
3.5              & Linearity                                                                   & \begin{tabular}[c]{@{}l@{}}Average Displacement = 39.66 $\pm$ 0.54 nm/mA; \\ Linearity = 99.4\% $\pm$ 0.33\%\end{tabular}                                                                                                                 \\ \hline
3.6              & Hysteresis                                                                  & Hyst = 2.10 $\pm$ 0.23\%                                                                                                                                                                                                                  \\ \hline
3.7              & Repeatability                                                               & \begin{tabular}[c]{@{}l@{}}Mirror shape is repeatable to below measurement precision\\ when mirror starts from rest;\\ %Some evidence was seen that repeatability could be affected by mirror\\ temperature or working-hours.
\end{tabular} \\ \hline
3.8              & Creep                                                                       & \begin{tabular}[c]{@{}l@{}}Measurements could not rule out creep $<$1000nm RMS, but \\ there is no indication of creep issues\end{tabular}                                                                                                                  \\ \hline
3.9              & Lifetime                                                                    & \begin{tabular}[c]{@{}l@{}}Over 110  million cycles have been completed \\ on each actuator with no change in performance\end{tabular}                                                                                                      \\ \hline
\end{tabular}               \end{table}

\subsection{Technology Development Plan}

\indent \indent TNO is in the process of constructing its first on-sky adaptive secondary mirror for use in the UH-88  telescope located on Maunakea.  This adaptive secondary mirror has an aspherical convex shape that is \diameter63cm.  It is fabricated using 210 actuators from the 2020 generation that can accommodate larger actuator spacing and be arranged in a circular grid.  This system is expected to be ready for lab testing in the summer of 2021 and be deployed in the telescope in early 2022. More information about the UH-88 ASM can be found in this conference's proceedings at SPIE No. 1144852 Chun et. al. 2020.

\begin{figure}[h!]
\centering
\begin{subfigure}{0.40\textwidth}
  \centering
  \includegraphics[width=1.0\linewidth]{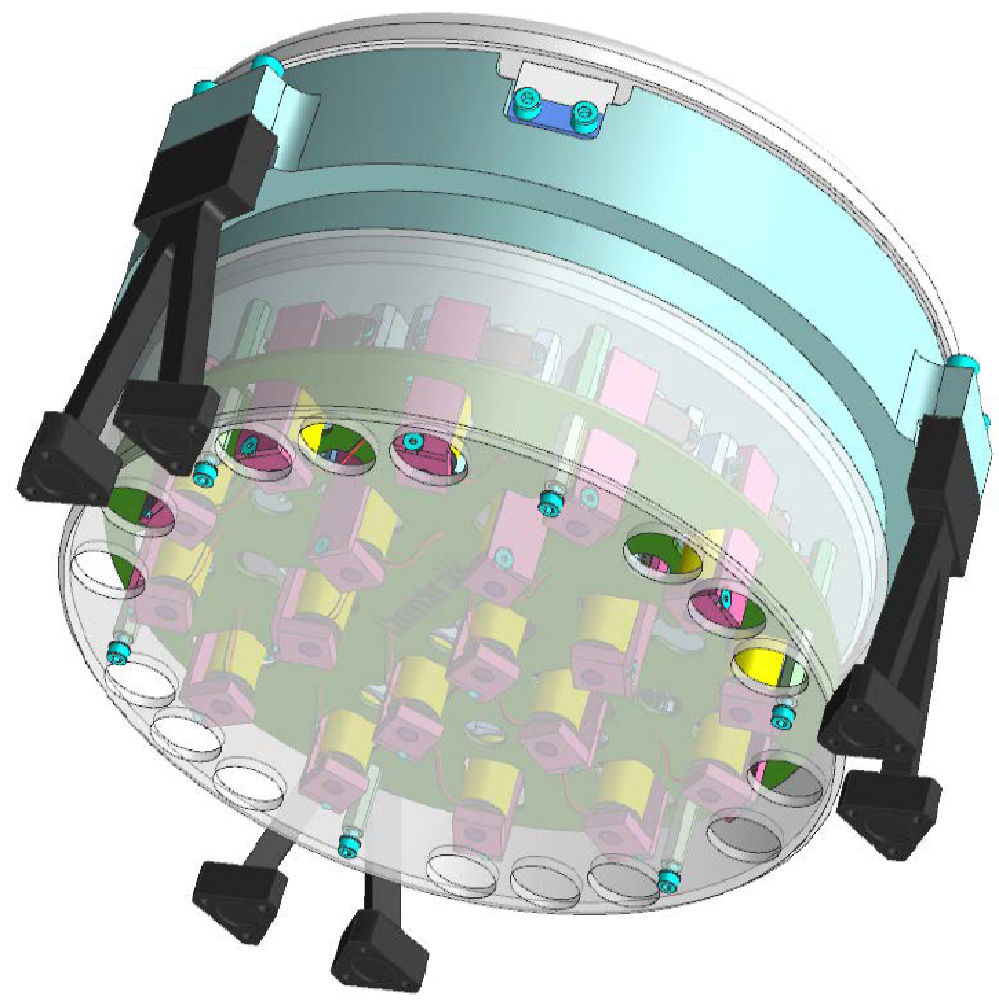}
\end{subfigure}
\begin{subfigure}{0.40\textwidth}
  \centering
   \includegraphics[width=1.0\linewidth]{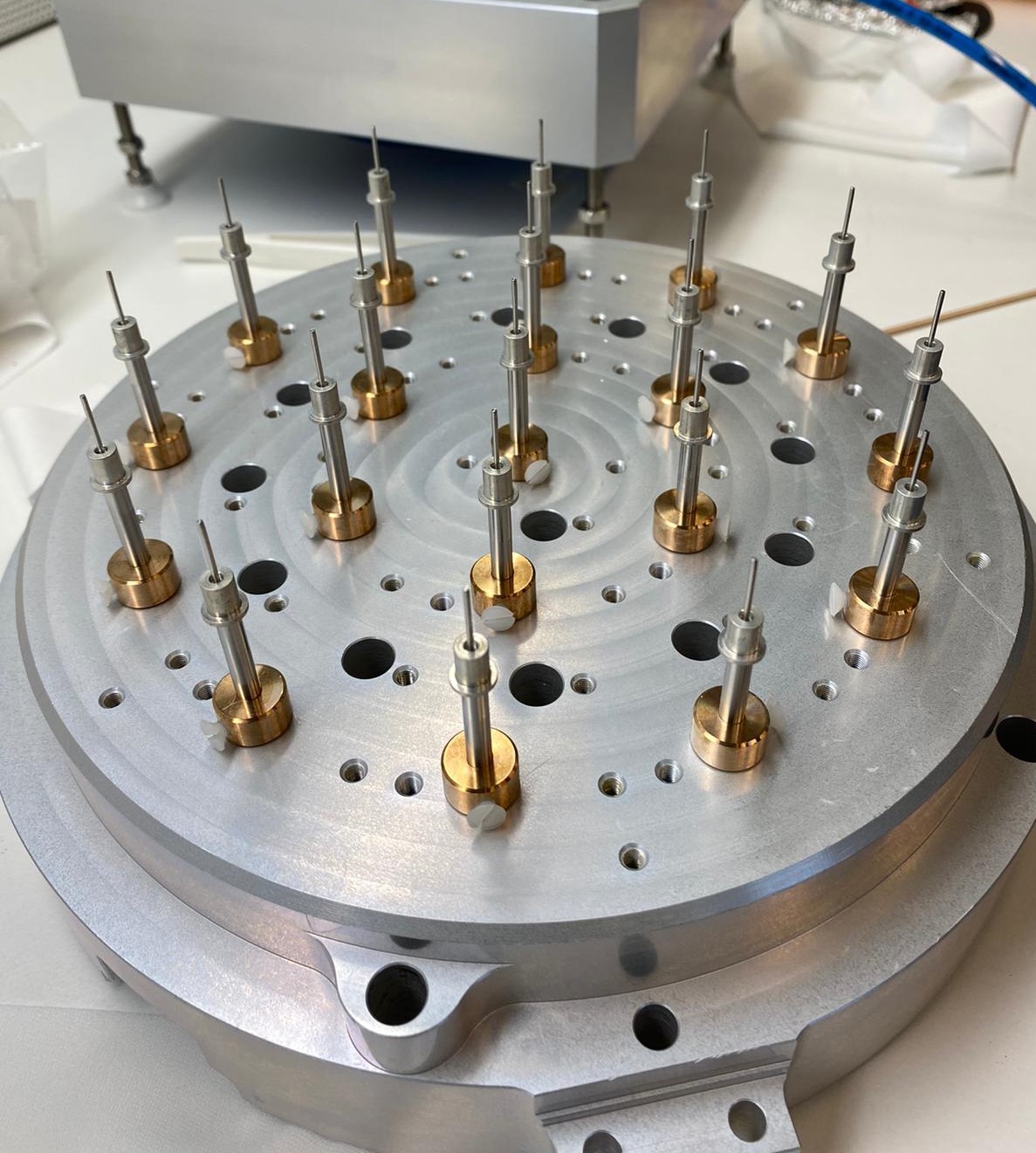}
 \end{subfigure}
 \caption[example]{\textbf{FLASH}: the First Lab Adaptive Secondary Holophote. The 19-actuator FLASH will be the first large-format deformable mirror designed  for use at the UCSC Lab for Adaptive Optics. It will be used to test the 2020 generation of actuators and explore  unexplained behaviors in the UH-88 ASM. (Left) CAD model of the FLASH. (Right) FLASH in the process of being constructed at TNO.  }
 \label{fig:FLASH}
   \end{figure}

TNO is developing two future lab-model prototypes for use at UCSC-LAO.  By December 2020, TNO will deliver the First Laboratory Adaptive Secondary Holophote (FLASH) (Figure \ref{fig:FLASH}). FLASH has a flat surface of \diameter160mm. It is constructed using 19 actuators from the 2020 version developed for the UH-88 ASM.  The testing presented in this paper will be repeated to verify the properties of the new actuators. Additional testing will also be performed to explore the effects of temperature, gravity, and any unexplained behaviors found in the UH-88 ASM.

Following the FLASH, the Second Laboratory Adaptive Secondary Holophote (SLASH) will be constructed using an aspherical convex facesheet. The SLASH will be \diameter370mm and contain 61 actuators from the next generation.   While the SLASH design is still being finalized, we are investigating the possibilities of integrating this prototype into the UC Lick Observatory’s Automated Planet Finder telescope as an adaptive secondary mirror. If this design is selected, SLASH may become the first on-sky ASM to be paired with a precision radial velocity instrument.

Both FLASH and SLASH will be constructed using thin facesheets formed using glass slumping techniques. This glass slumping is done in-house at the UCSC-LAO using flat borofloat sheets.  The glass is heated in a glass kiln and let sag into a fused silica mold over the course of several days.  It is then optically coated with aluminium to create a mirrored surface.   UCSC-LAO has produced the $\diameter$160mm flat facesheet for the FLASH using this technique in order to promote the technology readiness level of this fabrication approach. This technique will be developed to expand to meter sizes, greatly reducing the costs associated with producing flexible glass facesheets for use in ASMs for large telescopes. 

Design studies are in progress to understand the feasibility and constraints of building an adaptive secondary mirror for the UCO Lick Automated Planet Finder Telescope and the W.M. Keck Observatory with TNO technology. More information about these designs can be found in this conference's proceedings at SPIE No. 11448-233 Hinz et. al. 2020. 

\begin{table}[h!]
\caption{\label{tab:DMsplanned} \textbf{TNO Deformable Mirrors Tested by  UCSC-LAO}}
\begin{tabular}{|c|c|c|c|c|c|c|}
\hline
                                                                   & \textbf{\begin{tabular}[c]{@{}c@{}}\# of \\ Act.\end{tabular}} & \textbf{\begin{tabular}[c]{@{}c@{}}Diameter\\  (mm)\end{tabular}} & \textbf{\begin{tabular}[c]{@{}c@{}}Actuator\\ Type\end{tabular}}                     & \textbf{Facesheet}                                                                                        & \textbf{\begin{tabular}[c]{@{}c@{}}Communication\\  interface\end{tabular}} & \textbf{Status}                                                        \\ \hline
\textbf{DM3}                                                       & 57                                                             & 150                                                               & \begin{tabular}[c]{@{}c@{}}2016 model\\  designed for \\ lab prototypes\end{tabular} & \begin{tabular}[c]{@{}c@{}}standard flat wafer \\ (no optical figuring);\\  thickness = 1 mm\end{tabular} & Analog                                                                      & \begin{tabular}[c]{@{}c@{}}Testing\\ results \\ reported\end{tabular}  \\ \hline
\textbf{FLASH}                                                     & 19                                                             & 160                                                               & \begin{tabular}[c]{@{}c@{}}2020 model \\ designed for \\ UH-88\end{tabular}          & \begin{tabular}[c]{@{}c@{}}Flat slumped glass;\\ thickness = 3.3mm\end{tabular}                           & Analog                                                                      & \begin{tabular}[c]{@{}c@{}}To be \\ delivered \\ Dec 2020\end{tabular} \\ \hline
\textbf{\begin{tabular}[c]{@{}c@{}}SLASH\\ (pending)\end{tabular}} & 61                                                             & 400                                                               & \begin{tabular}[c]{@{}c@{}}2020 model \\ designed for \\ UH-88\end{tabular}          & \begin{tabular}[c]{@{}c@{}}Convex aspheric \\ slumped glass; \\ thickness = 3.3mm\end{tabular}            & Digital                                                                     & TBD                                                                    \\ \hline
\end{tabular}
\end{table}

\clearpage
% =========================================
\bibliographystyle{spiebib}   %>>>> makes bibtex use spiebib.bst
\bibliography{report}   %>>>> bibliography data in report.bib

\begin{thebibliography}{1}

\bibitem{Esposito2010}
{Esposito}, S., {Riccardi}, A., {Fini}, L., {Puglisi}, A.~T., {Pinna}, E.,
  {Xompero}, M., {Briguglio}, R., {Quir{\'o}s-Pacheco}, F., {Stefanini}, P.,
  {Guerra}, J.~C., {Busoni}, L., {Tozzi}, A., {Pieralli}, F., {Agapito}, G.,
  {Brusa-Zappellini}, G., {Demers}, R., {Brynnel}, J., {Arcidiacono}, C., and
  {Salinari}, P., ``{First light AO (FLAO) system for LBT: final integration,
  acceptance test in Europe, and preliminary on-sky commissioning results},''
  in [{\em Adaptive Optics Systems II}{\nolinebreak\hspace{0.1em}]},
  {Ellerbroek}, B.~L., {Hart}, M., {Hubin}, N., and {Wizinowich}, P.~L., eds.,
  {\em Society of Photo-Optical Instrumentation Engineers (SPIE) Conference
  Series} {\bf 7736},  773609 (July 2010).

\bibitem{Close2018}
{Close}, L.~M., {Males}, J.~R., {Morzinski}, K.~M., {Esposito}, S., {Riccardi},
  A., {Briguglio}, R., {Follette}, K.~B., {Wu}, Y.-L., {Pinna}, E., {Puglisi},
  A., {Xompero}, M., {Quiros}, F., and {Hinz}, P.~M., ``{Status of MagAO and
  review of astronomical science with visible light adaptive optics},'' in
  [{\em Adaptive Optics Systems VI}{\nolinebreak\hspace{0.1em}]},  {Close},
  L.~M., {Schreiber}, L., and {Schmidt}, D., eds., {\em Society of
  Photo-Optical Instrumentation Engineers (SPIE) Conference Series} {\bf
  10703},  107030L (July 2018).

\bibitem{Biasi2012}
{Biasi}, R., {Andrighettoni}, M., {Angerer}, G., {Mair}, C., {Pescoller}, D.,
  {Lazzarini}, P., {Anaclerio}, E., {Mantegazza}, M., {Gallieni}, D., {Vernet},
  E., {Arsenault}, R., {Madec}, P.~Y., {Duhoux}, P., {Riccardi}, A., {Xompero},
  M., {Briguglio}, R., {Manetti}, M., and {Morandini}, M., ``{VLT deformable
  secondary mirror: integration and electromechanical tests results},'' in
  [{\em Adaptive Optics Systems III}{\nolinebreak\hspace{0.1em}]},
  {Ellerbroek}, B.~L., {Marchetti}, E., and {V{\'e}ran}, J.-P., eds., {\em
  Society of Photo-Optical Instrumentation Engineers (SPIE) Conference Series}
  {\bf 8447},  84472G (July 2012).

\bibitem{Wildi2003}
{Wildi}, F.~P., {Brusa}, G., {Lloyd-Hart}, M., {Close}, L.~M., and {Riccardi},
  A., ``{First light of the 6.5-m MMT adaptive optics system},'' in [{\em
  Astronomical Adaptive Optics Systems and
  Applications}{\nolinebreak\hspace{0.1em}]},  {Tyson}, R.~K. and {Lloyd-Hart},
  M., eds., {\em Society of Photo-Optical Instrumentation Engineers (SPIE)
  Conference Series} {\bf 5169},  17--25 (Dec. 2003).

\bibitem{Hinz2016}
{Hinz}, P.~M., {Defr{\`e}re}, D., {Skemer}, A., {Bailey}, V., {Stone}, J.,
  {Spalding}, E., {Vaz}, A., {Pinna}, E., {Puglisi}, A., {Esposito}, S.,
  {Montoya}, M., {Downey}, E., {Leisenring}, J., {Durney}, O., {Hoffmann}, W.,
  {Hill}, J., {Millan-Gabet}, R., {Mennesson}, B., {Danchi}, W., {Morzinski},
  K., {Grenz}, P., {Skrutskie}, M., and {Ertel}, S., ``{Overview of LBTI: a
  multipurpose facility for high spatial resolution observations},'' in [{\em
  Optical and Infrared Interferometry and Imaging
  V}{\nolinebreak\hspace{0.1em}]},  {Malbet}, F., {Creech-Eakman}, M.~J., and
  {Tuthill}, P.~G., eds., {\em Society of Photo-Optical Instrumentation
  Engineers (SPIE) Conference Series} {\bf 9907},  990704 (Aug. 2016).

\bibitem{Kuiper2018}
{Kuiper}, S., {Doelman}, N., {Human}, J., {Saathof}, R., {Klop}, W., and
  {Maniscalco}, M., ``{Advances of TNO's electromagnetic deformable mirror
  development},'' in [{\em Advances in Optical and Mechanical Technologies for
  Telescopes and Instrumentation III}{\nolinebreak\hspace{0.1em}]},  {Navarro},
  R. and {Geyl}, R., eds., {\em Society of Photo-Optical Instrumentation
  Engineers (SPIE) Conference Series} {\bf 10706},  1070619 (July 2018).

\bibitem{Riccardi2003}
{Riccardi}, A., {Brusa}, G., {Salinari}, P., {Gallieni}, D., {Biasi}, R.,
  {Andrighettoni}, M., and {Martin}, H.~M., ``{Adaptive secondary mirrors for
  the Large Binocular Telescope},'' in [{\em Adaptive Optical System
  Technologies II}{\nolinebreak\hspace{0.1em}]},  {Wizinowich}, P.~L. and
  {Bonaccini}, D., eds., {\em Society of Photo-Optical Instrumentation
  Engineers (SPIE) Conference Series} {\bf 4839},  721--732 (Feb. 2003).

\end{thebibliography}

\end{document}